\documentclass[draft]{agujournal2019}
\usepackage{url} 
\usepackage{lineno}\let\linenumbers\relax
\usepackage{printlen}
\usepackage{amsmath}
\usepackage[LGRgreek]{mathastext}
\usepackage{xurl}
\usepackage[inline]{trackchanges} 
\usepackage{soul}
\usepackage[LGRgreek]{mathastext}
\nolinenumbers

\draftfalse

\journalname{Journal of Advances in Modeling Earth Systems}
\makeatletter\def\ps@headings{\def\@oddfoot{\centerline{\small --\the\c@page--}}\let\@evenfoot\@oddfoot\def\@oddhead{}\let\@evenhead\@oddhead}\makeatother
\begin{document}

\title{Benchmarking AIMIP models over the satellite era: Reproducing the training record does not imply physical consistency}

\authors{Ian Baxter\affil{1}, Katharine Rucker\affil{1}, Bingrui Wu\affil{2}, Pedram Hassanzadeh\affil{1}, Tiffany A. Shaw\affil{1}
}

\affiliation{1}{Department of the Geophysical Sciences, The University of Chicago, Chicago, IL, USA}
\affiliation{2}{School of Atmospheric Sciences, Nanjing University, Jiangsu, China}

\correspondingauthor{Ian Baxter}{itbaxter@uchicago.edu}

\begin{keypoints}
\item SST-forced AIMIP emulators vary widely in capturing satellite-era warming, with intermodel spread 2–6 times that of physics-based models
\item Emulators reproduce circulation trend patterns but underestimate regional amplitudes, even though the trends are in their training data
\item Most emulators cool the stratosphere without CO2 or ozone forcing, and one tracks the observed QBO only during its training period
\end{keypoints}

\begin{abstract}
AI-based weather and climate models (AIWCMs) from the AI Model Intercomparison Project (AIMIP), capable of stable decadal emulation, are evaluated against 1981–2014 temperature and circulation trends in ERA5 and physics-based simulations, and over their 2015–2024 testing period. The best AIWCMs capture land warming and the vertical structure of tropical warming comparably to physics-based models, while others produce little warming. This results in an AIWCM intermodel spread in thermodynamic trends 2–6 times that of physics-based models. AIWCMs reproduce the patterns of observed circulation trends, including the poleward shift of the austral summer jet, a European summer wave train trend, and the Matsuno–Gill response to tropical Pacific sea surface temperatures (SSTs), with comparable spread to physics-based models. Most AIWCMs shift the jet poleward consistently with their upper-level subtropical warming, as in reanalysis, though some do so without the accompanying temperature-gradient trend. However, they systematically underestimate the magnitudes of regional circulation trends, even with these trends present in their training data. Because ERA5 contains the full forced response, agreement with observed trends does not by itself imply physical consistency. Most AIWCMs cool the lower stratosphere despite lacking the forcing inputs that cause it. Only cBottle-1-3 reproduces the quasi-biennial oscillation, and only during its training period. A CO$_2$-ablation test shows that ACE2-ERA5 reproduces the thermodynamic and jet trends shown here through CO$_2$, whereas physics-based models primarily attribute them to SSTs. Reproducing the training record is therefore necessary but not sufficient for trusting AIWCMs outside it.
\end{abstract}

\section*{Plain Language Summary}
Physics-based climate models have reproduced many observed changes, but are slow and expensive. AI emulators learn how the atmosphere behaves from records of past weather and run hundreds of times faster. Before trusting them for climate projections, we need to know whether they reproduce observed changes for physically sensible reasons. We tested eight emulators, driven only by observed ocean surface temperatures and sea ice. The best capture warming over land and the tropical upper atmosphere compareably to physics-based models, but others barely warm, and emulators disagree far more than physics-based models do. They reproduce the patterns of observed wind changes, such as the jet stream's poleward shift or regional wind changes, but make some too weak. Because the emulators learned from a record that already contains these changes, matching it is not enough. Most cool the stratosphere with nothing in their inputs to cause it, some shift the jet stream without the temperature changes that normally accompany it, and one reproduces a regular stratospheric wind cycle only during the years it was trained on. These examples show that emulators can get the right answer for the wrong reason, so they need tests of physical consistency, not just comparison with observations.

\section{Introduction}
Recently, AI weather and climate models (AIWCMs) capable of multi-decadal stable emulation when forced with observed sea-surface temperature (SST), sea ice concentration (SIC), and, in some cases, carbon dioxide (CO$_2$) have emerged as promising complements to physics-based models~\cite{chapmanCAMulatorFastEmulation2025, kochkovNeuralGeneralCirculation2024, watt-meyerACE2AccuratelyLearning2025}. Most of these models build on the methodology of autoregressive AI weather emulators~\cite{biAccurateMediumrangeGlobal2023, pathakFourCastNetGlobalDatadriven2022, bouallegueRiseDataDrivenWeather2024, lamLearningSkillfulMediumrange2023a}, training on fast-timescale processes from ERA5 reanalysis data. 
Recent efforts have begun to expand the breadth of neural network architectures and approaches being used to emulate the atmosphere on climate timescales. This has resulted in the formation of the AI Model Intercomparison Project~\cite<AIMIP;>{hennAIMIPPhase12026}.
AIMIP has been designed to facilitate systematic comparison of climate trends in multi-decadal AMIP simulations by AI-driven models and to enhance the visibility and credibility of AIWCMs by learning about their current strengths and weaknesses. 

Over several generations, the understanding of the uncertainties across physics-based AMIP simulations has been refined into irreducible aleatoric, structural or epistemic, and scenario uncertainty~\cite{gatesAMIPAtmosphericModel1992,hawkinsPotentialNarrowUncertainty2009, massonClimateModelGenealogy2011, knuttiClimateModelGenealogy2013, kumaClimateModelCode2023}. 
Compared to physics-based models, AIWCMs are not built upon the same governing equations and do not share long parameterization or component ancestries, ~\cite{knuttiClimateModelGenealogy2013, kumaClimateModelCode2023} introducing new forms of uncertainty. 
Recent studies have focused on structural (epistemic) uncertainty or error in weather-focused AI models, including approximation, optimization, and generalization error~\cite{shalev-shwartzUnderstandingMachineLearning2014, luLearningNonlinearOperators2021, sunPredictingRegionalGray2026}. 
A shared training dataset, coarse-grained ERA5, removes training data differences, leaving architecture (approximation) and training (optimization) as the primary sources of intermodel spread~\cite{selzCanArtificialIntelligenceBased2023,scherEnsembleMethodsNeural2021,mahajanStatisticalEquivalenceAI2026,hassanzadehMissingButterflyPredicting2026}.
It is currently unclear how these remaining forms of uncertainty manifest as intermodel spread in AIWCMs and what should be expected of their longer-term behavior beyond weather timescales. Earlier studies focused primarily on ACE2-ERA5 and NeuralGCM~\cite{watt-meyerACE2AccuratelyLearning2025,kochkovNeuralGeneralCirculation2024, yuvalNeuralGeneralCirculation2026,baxterBenchmarkingAtmosphericCirculation2026,ruckerBenchmarkingRegionalThermodynamic2026, chenHierarchicalTestingHybrid2026,zhangEquilibriumResponseAtmospheric2026}.

In this study, we first characterize the spread across models in their regional thermodynamic and dynamic trends over the satellite era. This period provides the only global, observationally constrained record of the multi-decadal forced response of the atmosphere that can be tested directly against observations. Benchmarking the AIWCMs' ability to capture these observed trends in response to SST forcing is an important test of their skill for this boundary-value problem, with models shown to struggle with regional trends~\cite{simpsonConfrontingEarthSystem2025a,shawRegionalClimateChange2024a}. Because the AIWCMs are trained on ERA5 reanalysis through much of the satellite era (1979--2014), with 2015--2024 held out for testing, this benchmark is an ``in-sample'' and ``in-distribution'' evaluation (see Methods). The test is not trivial because the emulators are trained on globally averaged loss functions on short timesteps (hours to months).~\citeA{ruckerBenchmarkingRegionalThermodynamic2026} found that capturing the relatively small forced signal in regional thermodynamic trends is not guaranteed in ACE2-ERA5 or NeuralGCM. 

For AIWCMs to be useful for addressing questions currently addressed by physics-based models, it is important to establish whether they maintain the same physical relationships between variables, forcings, and processes. We define this physical consistency as the internal coherence between a model's temperature, humidity, and circulation responses to boundary forcings according to established dynamical theory and physics-based model experiments. For example, poleward transient eddy heat flux and wave breaking link meridional temperature gradient trends to Southern Hemisphere jet stream shifts~\cite{woollingsTrendsAtmosphericJet2023a}. 
Additionally, CO$_2$ forcing has been shown to drive stratospheric cooling along with tropospheric warming~\cite{manabeEffectsDoublingCO21975a,santerModeledObservedStratospheric2026}.
It is important that these changes are accompanied by the right mechanisms, as quantitative agreement with ERA5 does not mean an AIWCM adheres to the same mechanistic pathways as the reanalysis or physics-based models~\cite{simpsonConfrontingEarthSystem2025a}. Because CO$_2$, aerosol and ozone forcing co-vary with SSTs over the training period, an AIMIP model trained without them can assign their effects to the SSTs, an omitted-variable bias that yields the right trend for the wrong reason. In contrast, an emulator given CO$_2$ can use it as a shortcut to reproduce a trend~\cite{geirhosShortcutLearningDeep2020, brenowitzSpatiallyExtendedTests2019,ruckerBenchmarkingRegionalThermodynamic2026,clarkDisentanglingEffectsSea2026b}. Following the forcing partitioning of AMIP-piForcing simulations~\cite{andrewsEffectHistoricalSST2022a}, we therefore run ACE2-ERA5 with time-varying and with fixed CO$_2$ to separate the trends it produces through the SSTs from those it produces through the CO$_2$ input. A third possibility is that an emulator reproduces the training record itself rather than the response to its inputs, which can be tested by comparing its behavior within and after the training period.

In this study, we evaluate how well AIWCMs capture satellite-era trends in temperature and circulation compared with physics-based models. Performance is not solely based on agreement with ERA5, but also requires physical consistency with other variables or forcings. We also examine the spreads across each ensemble. Section 2 provides details on the protocols and models involved in AIMIP and AMIP as well as the metrics used to evaluate the models. In Section 3.1, we evaluate the ability of AIWCMs to represent the amplified surface temperature warming over land. We also show the vertical profiles of tropospheric temperature trends, and how they relate to low-level humidity trends and what AIWCMs ``learn'' from their training data. Section 3.2 examines the austral summer jet shift and its relationship with meridional gradients of temperature trends, stratospheric cooling and the quasi-biennial oscillation (QBO) across the training and 2015–2024 testing periods, and the underestimation of regional circulation trends that have been attributed to boundary forcings (e.g., aerosols, CO$_2$, SST).

\section{Materials and Methods} 

\subsection{AI Model Intercomparison Project (AIMIP)}
AIMIP is intended to document the current state of the art for all atmospheric models that include some significant AI component and are trained on ERA5 historical observations from 1979 to 2014~\cite{hennAIMIPPhase12026}. Model inferences are run from late 1978 to 2024, which includes a 3-month spin-up period and a 10-year test period (2015--2024). AIMIP protocols follow the Atmospheric Model Intercomparison Project (AMIP) as closely as possible, with the key exception that anthropogenic forcings (CO$_2$ and aerosols) are excluded from training and inference~\cite{gatesAMIPAtmosphericModel1992}. All AIMIP outputs examined in this study use monthly mean outputs at $1^{\circ}$ horizontal resolution. Six variables (ta, va, ua, hus, tas, huss) at seven or more vertical levels (1000 hPa, 850 hPa, 700 hPa, 500 hPa, 250 hPa, 100 hPa, 50 hPa) are examined for all models that provide them. The AIWCMs from AIMIP are trained and initialized with ERA5 reanalysis and forced with ERA5 monthly mean sea surface temperatures and sea ice concentrations linearly interpolated to model timesteps~\cite{hersbachERA5GlobalReanalysis2020}. Thus, ERA5 serves as our ground truth for the evaluation of satellite-era trends. See~\citeA{hennAIMIPPhase12026} for details of the individual models included in AIMIP.

NeuralGCM is a hybrid model that uses a dynamical core to simulate large-scale atmosphere dynamics and a learned neural network to represent the unresolved subgrid-scale tendencies \cite{kochkovNeuralGeneralCirculation2024, yuvalNeuralGeneralCirculation2026}. The neural network has been trained on the residual between the output of the dynamical core and the ERA5 dataset during 1979-2017 on 6-hour up to 5-day rollouts, and validated on 2018, leaving 2019-2023 as the out-of-sample test period. SST, SIC, and incident solar radiation are prescribed from the ERA5 dataset. There is no specific land component, though the model does include a surface embedding that takes the lowest-level atmospheric temperature and moisture as inputs. We run 37 ensemble members, each initialized at 10-day intervals throughout 1980 and run until 2023, using the deterministic model with 2.8° horizontal resolution and 37 pressure levels. To maintain simulation stability, the global mean surface pressure is fixed following \citeA{yuvalNeuralGeneralCirculation2026}.

ACE2-ERA5 is a fully data-driven model trained with a global mean-squared-error loss. We use ACE2-ERA5, which is trained on $1^{\circ}$ ERA5 data from 1940-1995, 2011-2019, and 2021-2022, and validated on 1996-2000, leaving 2001-2010 as the out-of-sample, in-distribution test period. SST, SIC, global-mean atmospheric CO$_2$, and downward shortwave radiative flux at the top of the atmosphere are prescribed from the ERA5 data provided in the ACE2-ERA5 repository. We run 37 ensemble members, in the same way as described for NeuralGCM, and integrate through 2022. ACE2-ERA5 is trained on ERA5 data vertically aggregated into eight hybrid sigma–pressure (terrain-following) layers and produces output on the same levels. Therefore, all ACE2-ERA5 data have been interpolated to pressure levels using the NCAR Command Language function vinth2p\_ecmwf.

A key decision in the AIMIP experiment design was the exclusion of CO$_2$. Therefore, to quantify the importance of this forcing in the 37-member ensemble from ACE2-ERA5, we run an identical 37-member ensemble forced with observed SST (as before) but fixed CO$_2$ averaged over the training years (1940-1995, 2011-2019, and 2021-2022); this is referred to as the ``fixed CO$_2$'' experiment hereafter. These two ACE2-ERA5 simulations will also be compared with ACE2-1-ERA5 from AIMIP, which was trained without CO$_2$.

\subsubsection{Physics-based models}
The physics-based land-atmosphere models used in this study are from the AMIP archive~\cite{gatesAMIPAtmosphericModel1992, eyringOverviewCoupledModel2016}. We use monthly mean 2m temperature and specific humidity, and temperature, specific humidity and horizontal winds ($u$, $v$) on standard pressure levels from 1981 to 2014. For eddy momentum flux convergence we use daily mean $u$ and $v$. A summary of the AMIP model outputs is available in Table S1. To quantify the role of SST forcing without time-evolving CO$_2$ or other forcings, we examine the same outputs from AMIP-piForcing simulations (Table S2;~\citeNP{andrewsEffectHistoricalSST2022a}). The AMIP-piForcing simulations, which are forced with historical SST but with external forcings fixed at the preindustrial level, are the closest comparisons to the AIMIP protocol, and their difference from AMIP is an analogue to the difference between the full forcing and fixed CO$_2$ ACE2-ERA5 experiments.

\subsection{Metrics}
Our analysis of physics-based models and AIWCMs focuses on the long-term changes in temperature and circulation represented by the linear trend from 1981 to 2014. For the reanalysis and for each physics-based model and AIWCM, the trend is quantified using ordinary least squares (OLS) linear regression. Absolute trends are reported in units per 34 yr. Spatial correlations are computed at $2.8^{\circ}$ resolution for comparison with NeuralGCM. Statistical significance is determined using a standard two-sided Student's t-test with an effective sample size. We define a model-observation discrepancy if the ERA5 trend lies outside the 5$^{th}$-95$^{th}$ percentile of each model distribution for large ensembles and the 5-member AIMIP models \cite{simpsonConfrontingEarthSystem2025a}. Five-member ensembles limit the within-model spread estimates and the significance tests. We report spread across the 37-member ensembles for NeuralGCM and ACE2-ERA5, and across the 5-member ensembles for the other AIMIP models. The 37-member NeuralGCM and ACE2-ERA5 ensembles are excluded from the pooled AIMIP range, and ranges and ratios without MD-1p5 are given in parentheses. Following general practice, intermodel spread is presented as the 5$^{th}$-95$^{th}$ percentile range of all realizations, comparing the AIMIP submissions with the AMIP and AMIP-piForcing members, in which models with more members carry more weight. 

\section{Results}

\subsection{Thermodynamic trends and variability}
\subsubsection{Amplified warming over land}

\begin{figure}
    \centering
    \includegraphics[width=1.0\linewidth]{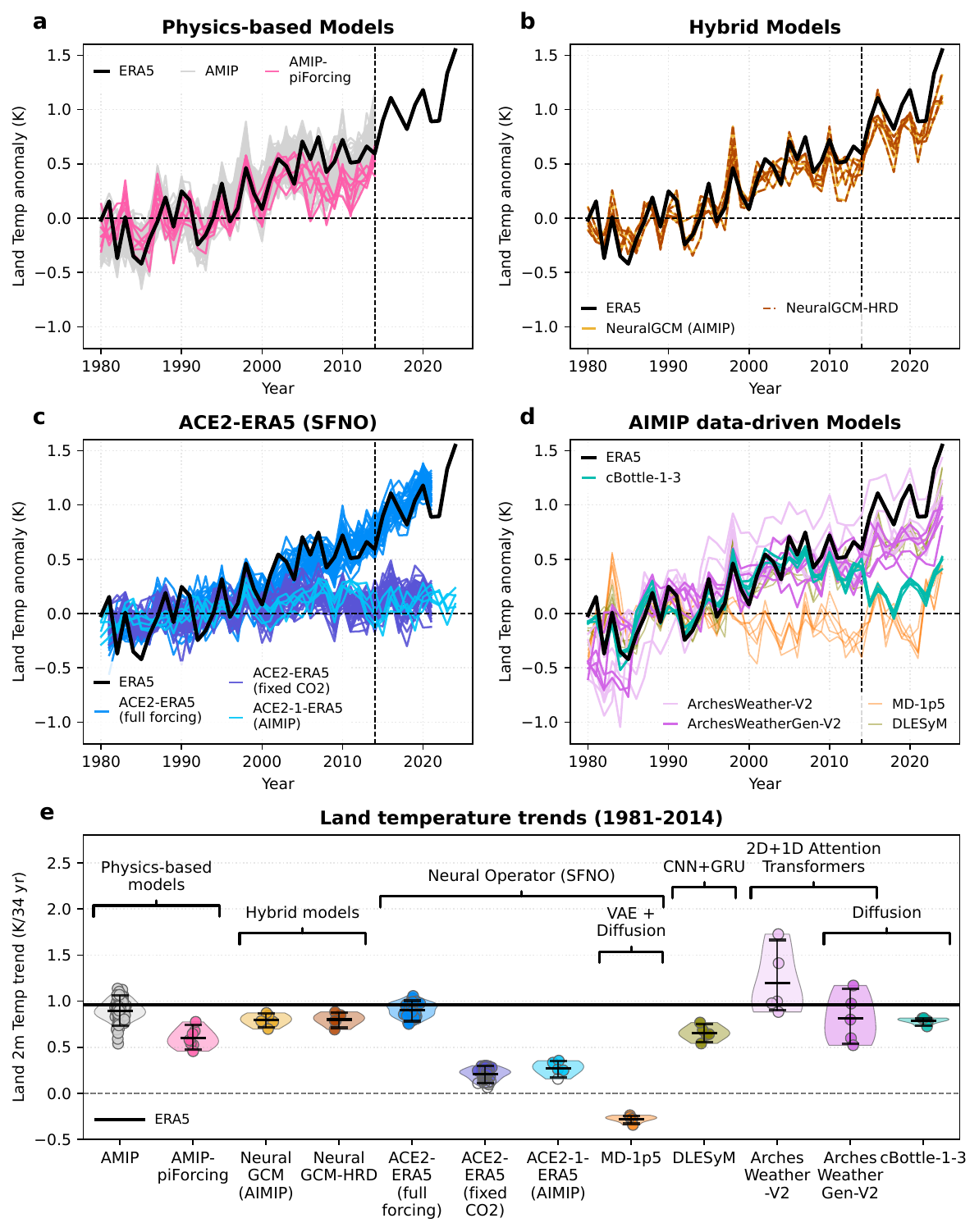}
    \caption{Anomalous annual mean near-surface (2m) temperatures over land (north of 60$^\circ$S) relative to 1981-1999 climatology from (a) AMIP (gray) and AMIP-piForcing (pink) models, (b) hybrid NeuralGCM models, (c) ACE2-ERA5 models, and (d) the remaining fully data-driven AIMIP models. (e) Linear trends from 1981 to 2014 of land near-surface temperature for each ensemble member (dots), with violins showing the distribution of members, black bars the ensemble mean, whiskers the $5^{th}-95^{th}$ percentiles and the horizontal black line ERA5. Filled dots indicate members with a statistically significant trend ($p<0.05$, two-sided t-test with an effective sample size from lag-1 autocorrelation); unfilled dots are not significant.}
    \label{fig:land_temperature_trends}
\end{figure}

We begin with the most prominent metric used for gauging the forced response in AIWCMs, amplified warming over land (north of 60$^{\circ}$S) relative to the ocean~\cite{zhangEquilibriumResponseAtmospheric2026}. Between 1981 and 2014, global mean temperature over land increased by $0.96$ K per 34 yr in reanalyses and has continued to increase to the present. Current physics-based climate models, especially those forced with observed SSTs from AMIP, are able to accurately capture the satellite-era warming from ERA5 over land (gray lines in Figure~\ref{fig:land_temperature_trends}a). Even with external forcings fixed at preindustrial levels, represented by the AMIP-piForcing simulations (pink in Figure~\ref{fig:land_temperature_trends}a), physics-based models still capture the reanalysis land warming up to the mid-2000s, and thereafter they exhibit slightly lower global mean land temperature between 2005 and 2014. This results in an ensemble mean trend in the AMIP-piForcing simulations about one-third weaker than in AMIP (0.60 vs. 0.90 K per 34 yr), with ERA5 lying outside the AMIP-piForcing 7-member spread (black line vs. second column in Figure~\ref{fig:land_temperature_trends}e). The qualitative consistency between AMIP and AMIP-piForcing suggests a physical connection between trends in SSTs and near-surface temperatures over land that can produce this amplified warming in response to only prescribed SSTs. 

The AIWCMs in AIMIP show a comparable ability to reproduce the land near-surface temperature time series over the recent four decades to that of the physics-based models (Figure~\ref{fig:land_temperature_trends}b-d). All realizations from the hybrid models, NeuralGCM and NeuralGCM-HRD, closely follow the ERA5 time series (Figure~\ref{fig:land_temperature_trends}b). ACE2-1-ERA5 shows very weak warming over land and MD-1p5 shows a weak cooling, while other AIWCMs show comparable trend magnitudes to the physics-based models (Figure~\ref{fig:land_temperature_trends}c,d). The ArchesWeather ensembles show larger spreads than the physics-based model ensembles or any other AIMIP model, coinciding with large, abrupt jumps in global land temperature that start in the stratosphere randomly between 1981 and 1994 (Figure~\ref{fig:land_temperature_trends}d). 

The overall AIMIP intermodel spread is also much larger than that of the physics-based models, with the 5$^{th}$-95$^{th}$ percentile range of the AIMIP ensemble spanning $-0.28$ to $1.18$ K per 34 yr ($0.26$ to $1.24$ K per 34 yr without MD-1p5), compared with $0.63$ to $1.06$ K per 34 yr for the physics-based models (Figure~\ref{fig:land_temperature_trends}e). The larger spread in the AIMIP ensemble may be related to differences in architecture, such as how each of the AIWCMs handles the land surface and the transfer of warming from SST to land~\cite{hennAIMIPPhase12026}. For example, DLESyM interpolates SSTs over the land points and underestimates land warming compared to ERA5. NeuralGCM does not impose a land surface representation but its response is consistent with its dynamical core, which transports SST warming to land temperature. ACE2-1-ERA5 predicts its own land surface temperature, whereas cBottle-1-3, MD-1p5, and ArchesWeather fill the SST forcing over land with a constant value~\cite{hennAIMIPPhase12026}. ACE2-1-ERA5, cBottle-1-3, and MD-1p5 fall furthest below ERA5 in the testing period, by 0.85 to 1.1 K in the 2015-2024 mean land temperature (Figure~\ref{fig:land_temperature_trends}c,d). 

The ratio of land-ocean warming links the near-surface land temperature trends to SST forcing (Figure S1). Across the AIWCMs, NeuralGCM, with its hybrid architecture, produces a response to prescribed SSTs most consistent with the physics-based models (gold and brown in Figure~\ref{fig:land_temperature_trends}b,e). We note that for almost all AIWCMs, trends continue into the testing period at similar rates. cBottle-1-3, across most metrics examined in this study, shows a magnitude similar to ERA5 over the AIMIP training period, but diverges back towards its climatological value in the testing period (Figure~\ref{fig:land_temperature_trends}d). Almost all the AIWCMs and the physics-based models exhibit amplified warming over land relative to the ocean, but underestimate this ratio compared with ERA5 (2.78, Figure S1). ArchesWeather-V2 and ArchesWeatherGen-V2 land/ocean trend ratios experience an abrupt jump between 1982 and 1994, where fitting a step at each member's jump year together with the trend lowers the 1981 to 2014 land/ocean trend ratio by $34\%$ in ArchesWeather-V2 and $19\%$ in ArchesWeatherGen-V2 (pink and magenta in Figure~\ref{fig:land_temperature_trends}d). DLESyM underestimates this land/ocean warming ratio, though it interpolates the SSTs over land grid points. ACE2-ERA5 no longer shows land warming when CO$_2$ is fixed, which is in contrast to the response seen in the physics-based models with fixed CO$_2$ (Figure~\ref{fig:land_temperature_trends}c vs.~\ref{fig:land_temperature_trends}a).

\subsubsection{Tropical upper-level warming}

\begin{figure}
    \centering
    \includegraphics[width=1.0\linewidth]{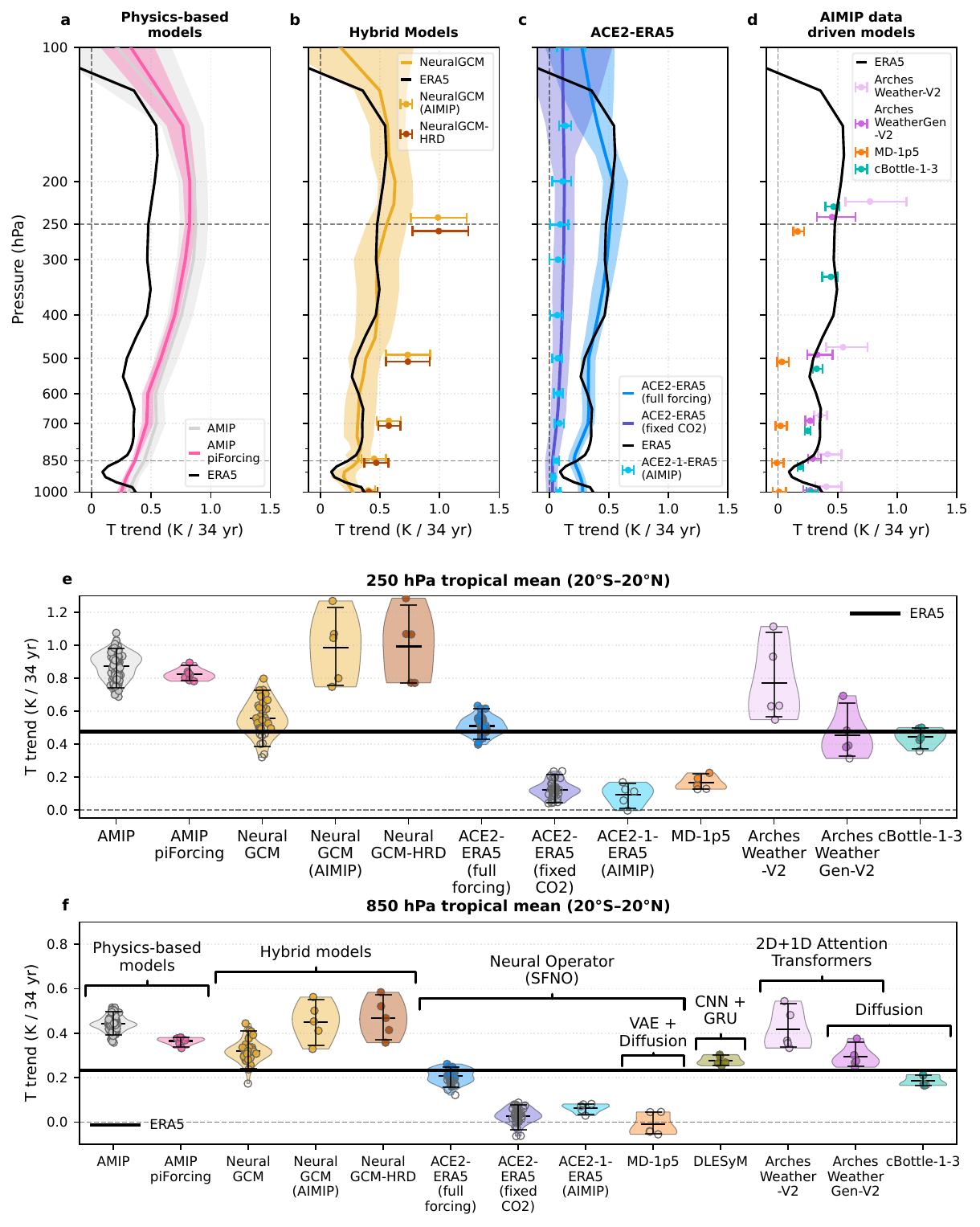}
    \caption{Linear trends from 1981 to 2014 in annual mean tropical mean (20$^\circ$S to 20$^\circ$N) temperature from (a) AMIP (gray) and AMIP-piForcing (pink) models, (b) NeuralGCM hybrid models, (c) ACE2-ERA5 models, and (d) the remaining data-driven AIMIP models. (e-f) Linear trends in annual mean tropical mean 250 and 850 hPa temperature. Horizontal thick black line shows ERA5 trend. Individual realizations shown by scatter points. Error bars show mean and 5$^{th}$-95$^{th}$ percentile bounds of each ensemble. Filled dots indicate members with a statistically significant trend according to a two-sided t-test.}
    \label{fig:tropical_trends}
\end{figure}

Another prominent signal of long-term climate trends is that of the warming of the tropical troposphere~\cite{manabeSensitivityGlobalClimate1980,po-chedleyInternalVariabilityForcing2022}. However, physics-based models overestimate the warming at upper levels by almost a factor of two, as shown by the gray line and shading in Figure~\ref{fig:tropical_trends}a. This stronger upper-level warming in physics-based models compared to observations and reanalyses has been linked to cloud and convective processes~\cite{mitchellRevisitingControversialIssue2013}, which are not explicitly represented in AIWCMs, but such subgrid-scale processes can be implicitly learned~\cite{hassanzadehMissingButterflyPredicting2026}. The role of these parameterizations and SST forcing in the overestimation is supported by a comparable magnitude of upper-tropospheric tropical temperature trends in the AMIP-piForcing simulations (pink in Figure~\ref{fig:tropical_trends}a). 

\citeA{kochkovNeuralGeneralCirculation2024} and \citeA{ruckerBenchmarkingRegionalThermodynamic2026} showed that some AIWCMs, NeuralGCM and ACE2-ERA5, can more accurately capture these tropical temperature trends compared with physics-based models. This is replicated in the vertical structure of tropical temperature trends in the 37-member NeuralGCM ensemble (gold in Figure~\ref{fig:tropical_trends}b). The 5-member NeuralGCM ensembles trained and run following the AIMIP protocols show stronger warming, especially at 250 hPa, that is more consistent with the physics-based models (gold and brown in Figure~\ref{fig:tropical_trends}b). We also show that ACE2-ERA5 with full forcing similarly captures the vertical temperature trends, but produces near-zero trends when CO$_2$ is fixed (Figure~\ref{fig:tropical_trends}c), highlighting that ACE2-ERA5 partitions the responses to SST and CO$_2$ opposite to that in physics-based models. 

The remaining AIWCMs show the amplified warming aloft and follow a similar relationship to ERA5 as that seen in their land temperature trends (Figure~\ref{fig:tropical_trends}d-f). MD-1p5 shows relatively weak upper-level temperature trends and near-zero trends at 850 hPa, comparable with ACE2-ERA5 fixed CO$_2$ and ACE2-1-ERA5. ArchesWeather-V2 uniformly overestimates warming, with the ERA5 trends lying outside its 5-member spread. ArchesWeatherGen-V2 captures the upper-level warming within its spread, but overestimates warming in the lower troposphere. cBottle-1-3 captures the upper-level warming trend from ERA5, but marginally underestimates the 850 hPa temperature trend. The AIWCMs also show much larger spreads in their tropical temperature trends compared to physics-based models, even within individual 5-member ensembles, where NeuralGCM (AIMIP) and ArchesWeather-V2 upper-tropospheric spreads are both $0.5$ K per 34 yr, compared with $0.2$ K per 34 yr for the physics-based models (Figure~\ref{fig:tropical_trends}e,f).
    
\begin{figure}
    \centering
    \includegraphics[width=1\linewidth]{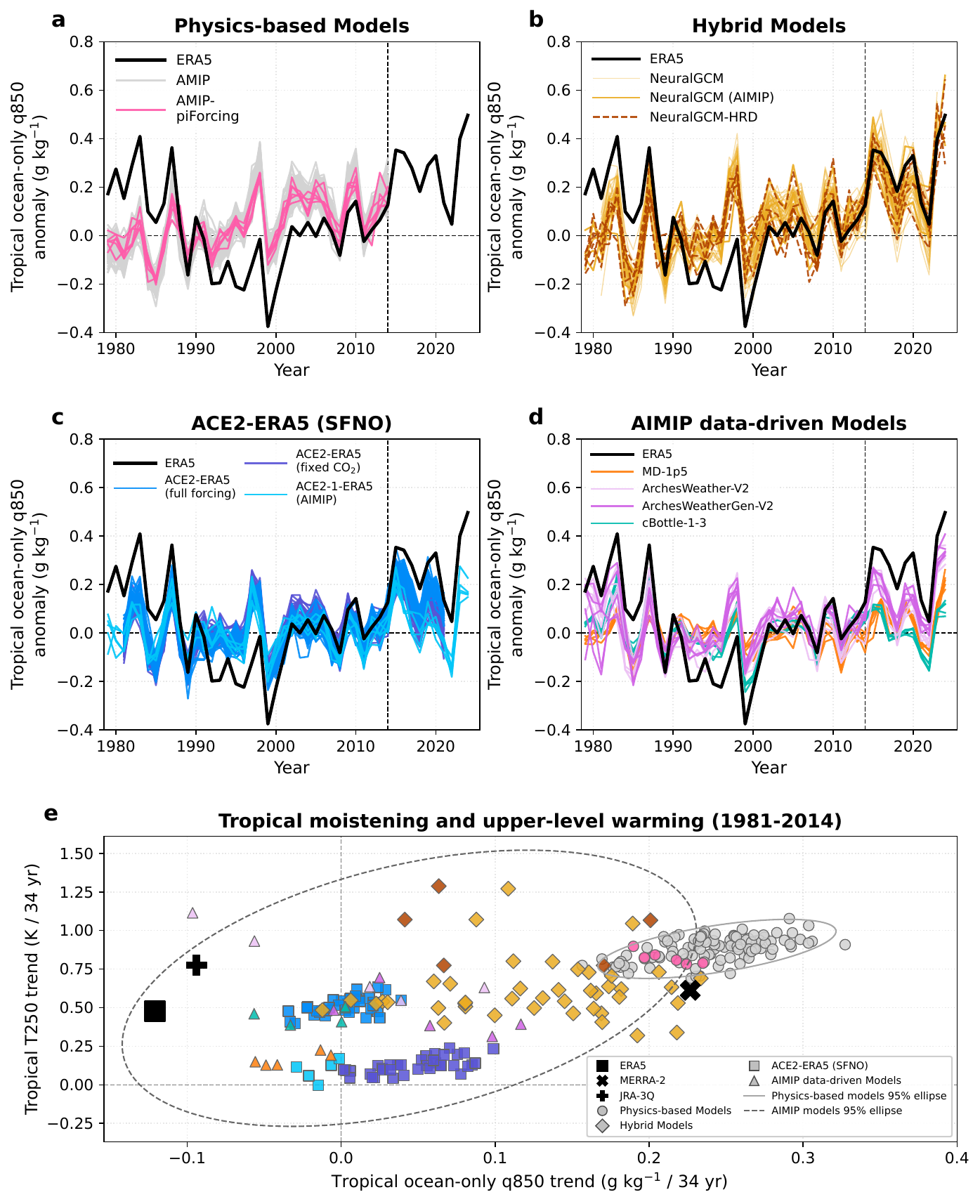}
    \caption{Tropical humidity and upper-level temperature trends. (a-d) Annual-mean tropical (20$^\circ$S-20$^\circ$N) ocean-only 850 hPa specific humidity (q850) anomalies relative to each member's 1981-1999 mean for (a) physics-based models, (b) hybrid models, (c) ACE2-ERA5 runs and (d) the other AIMIP data-driven models, with ERA5 in black. The dashed vertical line marks the end of the AIMIP training period (2014). (e) 1981-2014 trends of q850 against tropical 250 hPa temperature (T250) for each ensemble member, with ERA5, JRA-3Q and MERRA-2 as a black square, plus and cross. Marker shapes denote the model groups in (a-d). Ellipses enclose 95\% of the physics-based (solid) and AIMIP (dashed; 5-member submissions without MD-1p5) model members, with each model weighted equally.}
    \label{fig:zm_trends_hus}
\end{figure}

\subsubsection{Tropical humidity}

Tropical warming aloft is tied to low-level moisture, which fuels convection and the release of latent heat, which warms the higher levels of the atmosphere (Figure~\ref{fig:tropical_trends}). Therefore, it is possible that the overestimation of upper-level tropical warming seen in physics-based models compared with reanalyses is linked to differences in their low-level humidity trends (Figure~\ref{fig:zm_trends_hus}). ERA5 shows a decrease in annual-mean low-level specific humidity over the tropics at 850 hPa, followed by a comparable magnitude increase beginning in the late 1990s (black lines in Figure~\ref{fig:zm_trends_hus}a-d). In contrast, physics-based models show a steadily increasing humidity (Figure~\ref{fig:zm_trends_hus}a). This discrepancy is associated with more positive moistening trends in physics-based models throughout the tropics that dominate over the drying signal in the eastern tropical Pacific (Figures S2 and S3). The physics-based models show a strong relationship between low-level humidity and upper-tropospheric temperature trends ($r=0.60$) following moist adiabatic scaling~\cite{emanuelLargescaleCirculationsConvecting1994, santerAmplificationSurfaceTemperature2005, zhangProjectionsTropicalHeat2021, keilVariationsTropicalLapse2021}. This discrepancy may reflect both the reanalysis, whose tropical drying arises from a potentially spurious decrease before about 1993~\cite{allanGlobalChangesWater2022}, and the physics-based models, whose stronger moistening is consistent with enhanced convection and latent heating that warms their upper troposphere faster than in ERA5. 

AIWCMs replace these parameterizations with data-driven representations learned from coarse-grained data that represent the cumulative response of fine-scale processes on the larger scales~\cite{kochkovNeuralGeneralCirculation2024,watt-meyerACE2AccuratelyLearning2025,hassanzadehMissingButterflyPredicting2026}. However, NeuralGCM, which most directly replaces those parameterizations with a neural network, shows a weakly increasing tropical humidity trend, closer to that in the AMIP and AMIP-piForcing simulations than to ERA5 (Figure~\ref{fig:zm_trends_hus}b \& S3d,e). ArchesWeather-V2 and ArchesWeatherGen-V2 show large spreads in their tropical humidity trends that range from weakly positive like NeuralGCM to close to ERA5 (pink and magenta triangles in Figure~\ref{fig:zm_trends_hus}e). The other AIMIP models show trends closer to zero (Figure~\ref{fig:zm_trends_hus}). Although the AIMIP ellipse is 12 times larger in its spread (13 times without MD-1p5), the AIMIP members show the same positive relationship between low-level humidity and upper-tropospheric temperature trends as the physics-based members, though a weaker one ($r=0.50$, or $r=0.41$ without MD-1p5, vs. $r=0.60$, member correlations with each model weighted equally). ERA5 still lies at the border of the 95\% ellipse based on AIMIP (dashed gray ellipse in Figure~\ref{fig:zm_trends_hus}e). 

Following a consistent relationship, the physics-based models moisten and warm aloft more strongly than ERA5, while most AIWCMs moisten weakly and warm aloft at a similar rate to ERA5. Neither reproduces ERA5's humidity record, whose drying is not in agreement across reanalysis products (black square, plus and cross in Figure~\ref{fig:zm_trends_hus}e), despite the trend being present in the ERA5 training data. This mismatch between moisture and temperature highlights a challenging question of what AIWCMs learn from their training data and how these AIWCMs represent the large-scale response to finer-scale processes like convection and clouds on long timescales.  

\subsection{Circulation trends and variability}
\subsubsection{Zonal-mean circulation and its relationship with upper-level temperature}

\begin{figure}
    \centering
    \includegraphics[width=1\linewidth]{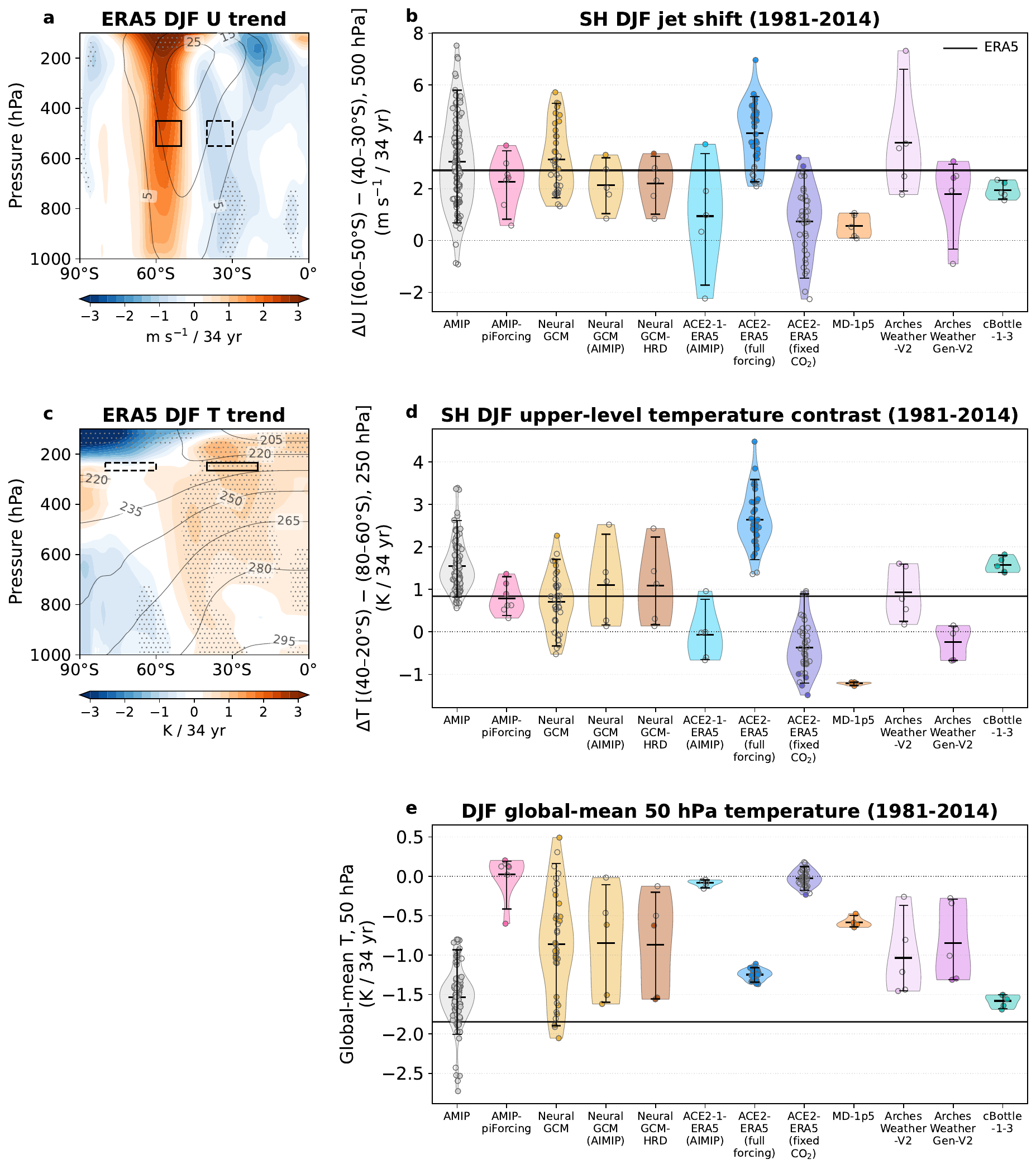}
    \caption{Southern Hemisphere DJF jet shift, upper-level temperature gradient and global stratospheric temperature trends. (a) ERA5 DJF zonal-mean zonal wind trend from 1981 to 2014 (shading) and 1981-2014 climatology (contours, m s$^{-1}$). (b) Trends in the jet-shift index, the cos(latitude)-weighted zonal-mean zonal wind averaged over 60-50$^\circ$S minus that over 40-30$^\circ$S at 500 hPa (solid and dashed boxes in a, respectively); positive values indicate a poleward shift or strengthening of the jet. (c) As in (a) but for zonal-mean temperature (contours in K). (d) Trends in the upper-level meridional temperature gradient, the zonal-mean temperature averaged over 40-20$^\circ$S minus that over 80-60$^\circ$S at 250 hPa (solid and dashed boxes in c); positive values indicate a strengthening of the equator-to-pole gradient. (e) Trends in the DJF global-mean (cos(latitude)-weighted) temperature at 50 hPa, the AIMIP output level that lies in the stratosphere at all latitudes. Stippling in (a) and (c) indicates significant ERA5 trends according to a two-sided t-test. In (b), (d) and (e), each dot is one ensemble member, violins show the distribution of members, black bars show the ensemble mean and whiskers the 5$^{th}$-95$^{th}$ percentiles, and the horizontal black line is ERA5. Filled dots indicate members with a statistically significant trend.}
    \label{fig:jet_shift_scatter}
\end{figure}

We next examine how temperature trends might be associated with extratropical circulation trends in AIWCMs.~\citeA{woollingsTrendsAtmosphericJet2023a} suggested that during December-January-February (DJF) in the Southern Hemisphere (SH), strengthened upper-level meridional temperature gradients, tropical warming and the lower stratospheric cooling over Antarctica are associated with a poleward shift of the zonal mean winds and jet streams through an eddy-mediated response. Over the satellite era, ERA5 has shown a poleward shift of the DJF SH jet streams, represented by a strengthening of the zonal mean winds poleward of the climatological maxima and weakening on the equatorward side ($2.71$ m s$^{-1}$, Figure~\ref{fig:jet_shift_scatter}a). Both AMIP and AMIP-piForcing simulations show a jet shift comparable to that in ERA5 ($3.03$ m s$^{-1}$ and $2.27$ m s$^{-1}$, gray and pink in Figure~\ref{fig:jet_shift_scatter}b). This poleward shift is accompanied by an enhanced equator-to-pole temperature gradient ($0.84$ K at 250 hPa, Figure~\ref{fig:jet_shift_scatter}c), consistent with a stronger tropical warming (Figure~\ref{fig:tropical_trends}e,~\ref{fig:jet_shift_scatter}d;~\citeNP{griseOzoneHoleIndirect2013},~\citeNP{diazDisentanglingAnthropogenicEffects2026}). However, the AMIP simulations exhibit much stronger cooling at 250 hPa near the pole compared to ERA5 ($-0.60$ K vs. $+0.05$ K over 80-60$^\circ$S), resulting in a much larger meridional temperature gradient trend ($1.55$ K). The AMIP models' strong lower stratospheric cooling over Antarctica is reduced in the AMIP-piForcing simulations. AMIP-piForcing simulations are closer to ERA5 because their meridional temperature gradient at 250 hPa primarily reflects subtropical warming ($0.79$ K, Figure~\ref{fig:jet_shift_scatter}d,e).

The AIWCMs, apart from cBottle-1-3 and MD-1p5, capture the poleward shift of the DJF SH jet within their spreads (Figure~\ref{fig:jet_shift_scatter}b; zonal-mean zonal wind trends for each ensemble in Figure S4). ACE2-ERA5 (full forcing) and ArchesWeather-V2 show a stronger poleward shift than ERA5 (Figure~\ref{fig:jet_shift_scatter}b). AIMIP also shows a smaller spread in trend magnitudes of the SH DJF jet shift across the ensemble compared with physics-based models (0.77 times their $5^{th}$-$95^{th}$ percentile range, 0.80 without MD-1p5). The AIWCMs are also able to capture tropical temperature trends comparably to the physics-based models at 250 hPa (Figure~\ref{fig:jet_shift_scatter}d), but struggle at higher altitudes within the stratosphere (Figure S5). 

The DJF SH poleward jet shift shows a positive relationship with the upper-level meridional temperature gradient ($r=0.68$ across AIWCM members, Figure~\ref{fig:jet_shift_scatter}b,d), consistent with the mechanism in~\citeA{woollingsTrendsAtmosphericJet2023a}. After removing AIWCMs that do not warm globally (ACE2-1-ERA5, MD-1p5 and ACE2-ERA5 with fixed CO$_2$), the correlation across AIWCMs remains significant, as in the physics-based models ($r=0.49$ vs. $r=0.39$). AIWCMs also show consistency in the shift of eddy momentum flux convergence towards higher latitudes and faster phase speeds that contribute to the poleward shift (\citeNP{chenPhaseSpeedSpectra2007a,hayashiGeneralizedMethodResolving1971,randelPhaseSpeedSpectra1991,lutskoLowerTroposphericEddyMomentum2017}; Figure S6). Despite this consistent relationship between winds and temperature, and their ability to capture the magnitude of the jet-shift trend within their spreads, some AIWCMs, ACE2-1-ERA5 and ArchesWeatherGen-V2, produce a jet shift without an increased meridional upper-level temperature-gradient trend (Figure~\ref{fig:jet_shift_scatter}d). The AIWCMs are likely capturing the poleward jet shift through subtropical warming, as proposed in~\citeA{woollingsTrendsAtmosphericJet2023a}, while physics-based models and AIWCMs trained with CO$_2$ show much stronger cooling at 250 hPa at high latitudes relative to the magnitude of their jet shift.

Though the AIWCMs without 250 hPa cooling over Antarctica are most consistent with ERA5, almost all exhibit a broad cooling across most of the stratosphere (Figure~\ref{fig:jet_shift_scatter}e). Previous studies, along with the physics-based models shown here (Figure~\ref{fig:jet_shift_scatter}e, S5b-c), suggest that CO$_2$ and ozone have cooled the stratosphere~\cite{manabeEffectsDoublingCO21975a, wuAtmosphericCirculationResponse2012, santerModeledObservedStratospheric2026}. This is a case where the AIWCMs are not consistent with the forced response in physics-based models. One exception is ACE2-ERA5, where ACE2-1-ERA5 and ACE2-ERA5 with fixed CO$_2$ show near-zero stratospheric temperature trends. The AIMIP protocol also calls for a minimum of seven vertical levels, of which only two represent the stratosphere, and some AIWCMs do not train their emulators with the full stratosphere. This may not be sufficient to capture the important processes that have contributed to satellite-era stratospheric temperature trends. 

\begin{figure}
    \centering
    \includegraphics[width=1\linewidth,height=0.72\textheight,keepaspectratio]{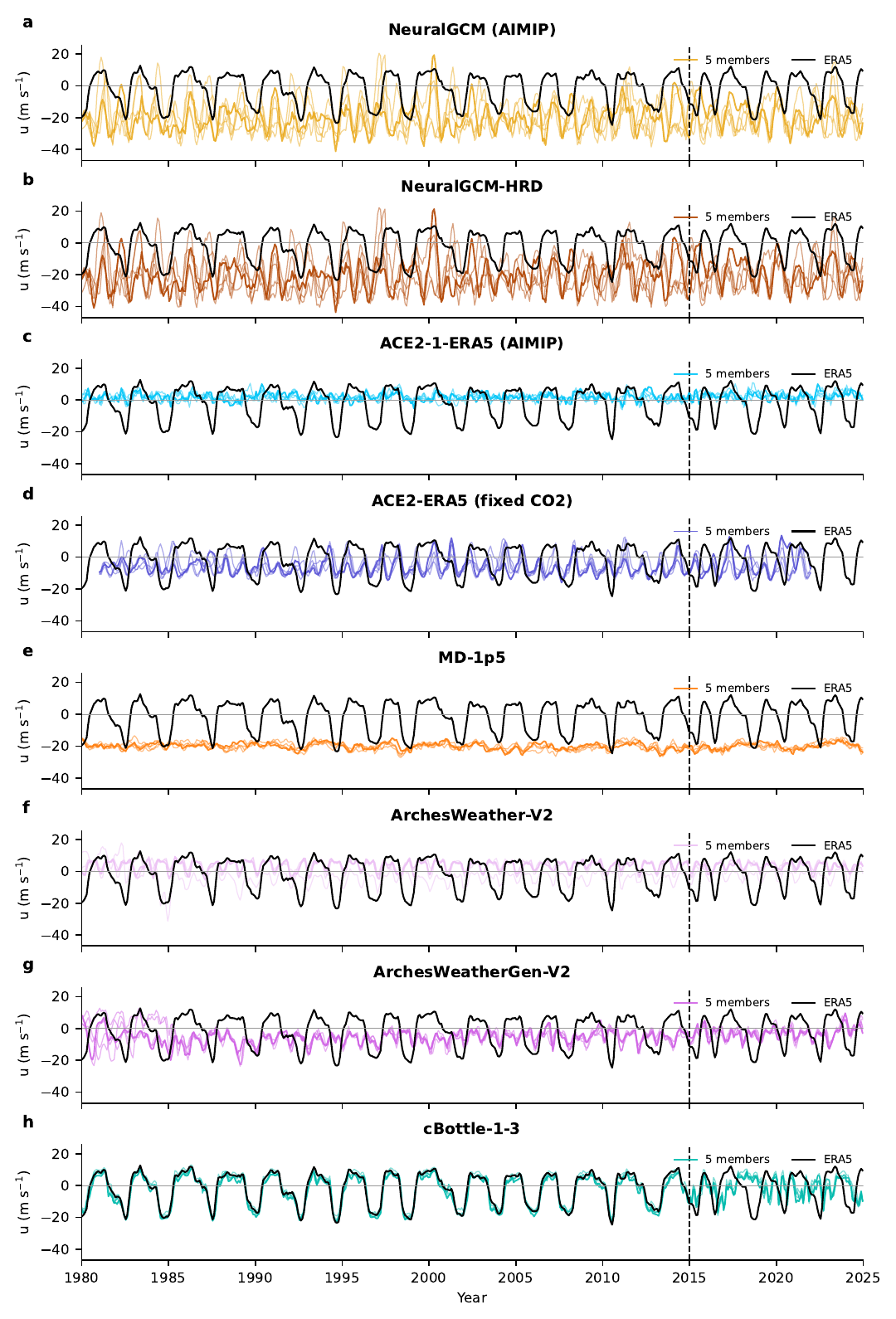}
    \caption{Equatorial stratospheric zonal wind from 1980 to 2024. Monthly zonal-mean zonal wind at 50 hPa averaged over
    10$^\circ$S-10$^\circ$N for (a-h) the members of each AIWCM (colored lines; for ACE2-ERA5 fixed CO$_2$, five members from 1981 to 2021) and ERA5 (black). The dashed line marks the
    end of the AIMIP training period (1979-2014).}
    \label{fig:qbo_u50}
\end{figure}

Low vertical resolution may also explain the AIWCMs’ inability to produce regular oscillations in their equatorial stratospheric winds, like the QBO (\citeNP{baxterBenchmarkingAtmosphericCirculation2026}; Figure~\ref{fig:qbo_u50}). cBottle-1-3 reproduces the reanalysis QBO from 1980 to 2014 ($r=0.98$ with ERA5 in every member), within the AIMIP training period, but not from 2015 to 2024 ($r=-0.1$). The other AIWCMs show either weak equatorial wind variability (ACE2-1-ERA5, ArchesWeather-V2 and MD-1p5) or large variability without a regular period (NeuralGCM from AIMIP and ACE2-ERA5 fixed CO$_2$). Because the QBO is generated internally rather than forced by SSTs, it provides a clean test: reproducing it during the training period but not afterward shows that agreement over the training period can arise without the underlying mechanism (see Section 4.2). 

\subsubsection{Regional circulation}

\begin{figure}
    \centering
    \includegraphics[width=1\linewidth]{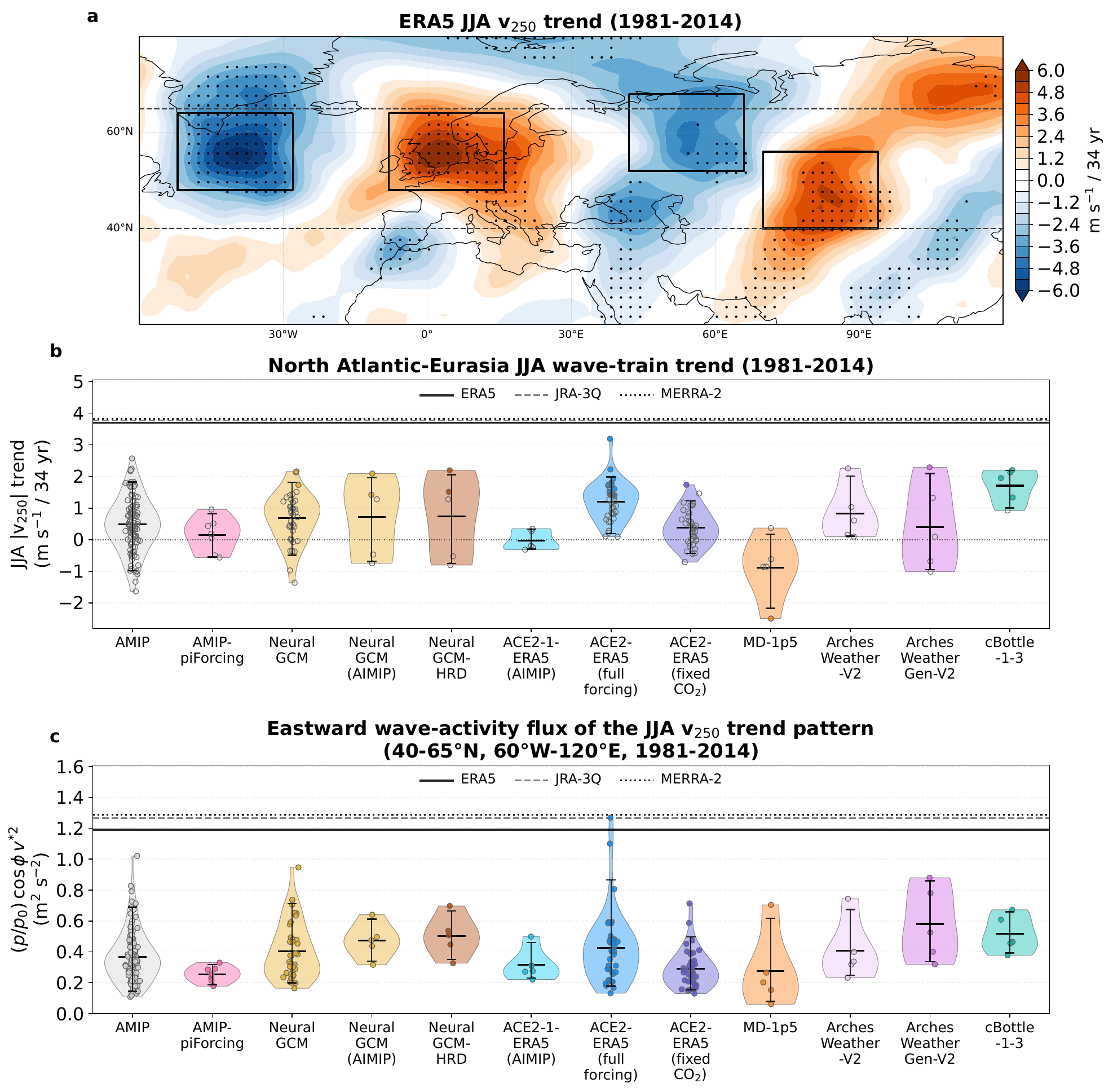}
    \caption{Boreal summer (JJA) 250 hPa meridional wind trends over the North Atlantic and Eurasia, 1981-2014. (a) ERA5 trend. Black dots indicate local significance according to a two-sided Student's t-test. The four boxes mark the main centers of the ERA5 wave train. (b) Wave-train index for each ensemble member, defined as the latitude-weighted mean trend in each box multiplied by the sign of the ERA5 trend in that box and averaged over the four boxes. Positive values indicate a wave train in phase with ERA5 and negative values a reversed one. (c) Eastward wave-activity flux of each member's trend pattern, $(p/p_0)\cos\phi\,v^{*2}$ averaged over 40-65$^\circ$N, 60$^\circ$W-120$^\circ$E (dashed lines in a), where $v^*$ is the trend minus its zonal mean~\cite{plumbThreeDimensionalPropagationStationary1985}. In (b) and (c), violins, black bars and whiskers are as in Figure~\ref{fig:jet_shift_scatter}, and the horizontal lines show the trends in ERA5 (solid black), JRA-3Q (dashed gray;~\citeNP{kosakaJRA3QReanalysis2024}) and MERRA-2 (dotted black;~\citeNP{gelaroModernEraRetrospectiveAnalysis2017}). Filled dots in (b) indicate members with a statistically significant trend according to a two-sided t-test.}
    \label{fig:europe_vwinds_trends}
\end{figure}

The AIWCMs are trained to minimize a global loss, which may hide or redistribute error across localized, regional trends. Therefore, we also examine European and central Pacific circulation changes identified over the historical record. In summertime, physics-based models and AIWCMs struggle to reproduce the amplitude of the wave train pattern in meridional wind trends over the North Atlantic and Europe (Figure~\ref{fig:europe_vwinds_trends}). We define two simple metrics to evaluate European meridional wind trend performance. The first is an area-weighted mean over the four primary action centers chosen based on ERA5 (black boxes in Figure~\ref{fig:europe_vwinds_trends}a, trends in Figure~\ref{fig:europe_vwinds_trends}b). We also show the trend in eastward wave-activity flux, proportional to the square of the JJA eddy meridional wind trend (zonal mean removed), averaged between 40$^\circ$N and 65$^\circ$N (dashed lines in Figure~\ref{fig:europe_vwinds_trends}a, trends in Figure~\ref{fig:europe_vwinds_trends}c). The index based on action centers mixes amplitude and position of the wave train trends, while the wave-activity flux measures only the amplitude of the wind trends over the North Atlantic and Europe. 

Physics-based models underestimate the trends in both meridional wind indices (Figure~\ref{fig:europe_vwinds_trends}b,c). Yet they produce statistically significant spatial correlations with ERA5, generating a similarly positioned wave train pattern as that found in ERA5 (Figure~S7b,c, $r=0.56$ and $r=0.39$). Thus, the physics-based models struggle primarily with capturing the amplitude of the meridional wind trends, not their position. It has been suggested that this may be due to an underestimation of the response to anthropogenic aerosol emissions~\cite{kangSummertimeStationaryWave}. Since AIWCMs have the observed forced response present in their training data, they might be expected to capture the magnitude of these trends despite not including aerosols as an input.

AIWCMs show the same underestimation of the meridional wind trends over Europe as the physics-based models (Figure~\ref{fig:europe_vwinds_trends}b,c). Pooling all members, the 5$^{th}$-95$^{th}$ percentile range of the wave-train index is $-0.91$ to $2.22$ m s$^{-1}$ per 34 yr for the AIMIP models ($-0.78$ to $2.23$ without MD-1p5) and $-0.96$ to $1.81$ m s$^{-1}$ per 34 yr for the physics-based models, and the corresponding ranges of the eastward wave-activity flux are $0.19$ to $0.75$ ($0.25$ to $0.76$) and $0.15$ to $0.68$ m$^2$ s$^{-2}$ per 34 yr. ERA5 ($3.70$ m s$^{-1}$ and $1.19$ m$^2$ s$^{-2}$ per 34 yr) lies above every member of every ensemble for the wave-train index, and only one ACE2-ERA5 (full forcing) member exceeds it for the flux. Unlike the thermodynamic trends, the spread of these trends across AIMIP is therefore comparable to that across physics-based models. As in the physics-based models, the deficit is mainly in amplitude rather than position. Because the wave-activity flux does not depend on the phase of the wave train, its underestimation in every ensemble indicates a wave train that is too weak rather than misplaced. Although the AIWCMs can roughly capture the overall pattern and members with the highest pattern correlations can reproduce significant spatial correlations with the ERA5 wave train (Figure~S8), none reaches its amplitude. The presence of the full forced response in the training data is therefore not sufficient for the AIWCMs to reproduce it. In contrast to the temperature trends, ACE2-ERA5 partitions this signal similarly to the physics-based models, as fixing CO$_2$ retains 31\% of the wave-train index and 68\% of the wave-activity flux trend, compared with 31\% and 69\% in AMIP-piForcing relative to AMIP.

\begin{figure}
    \centering
    \includegraphics[width=1\linewidth,height=0.62\textheight,keepaspectratio]{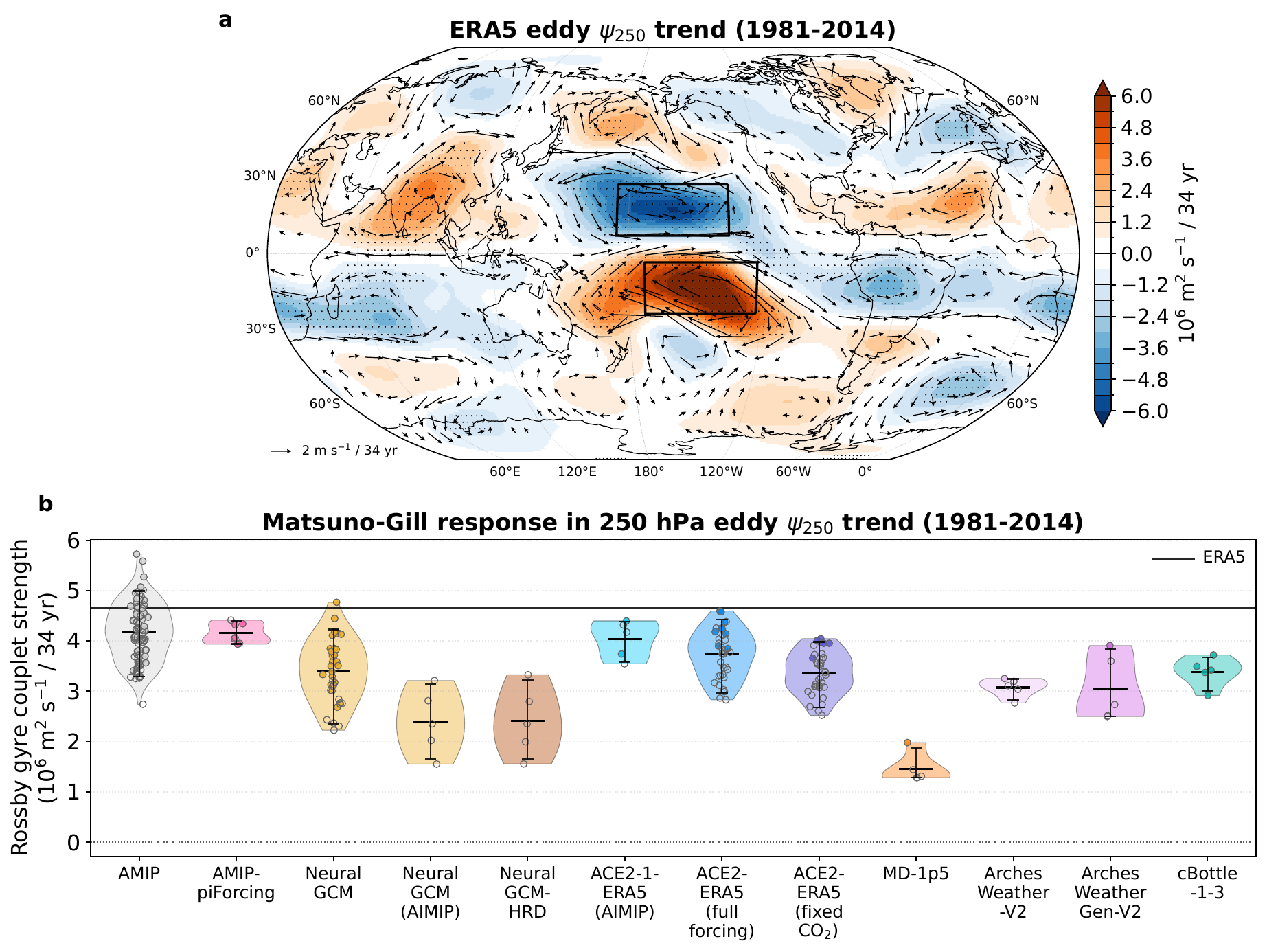}
    \caption{The Matsuno-Gill response in 1981-2014 trends of the 250 hPa eddy stream function (zonal mean removed). (a) ERA5 trend. Stippling marks where the trend is statistically significant according to a two-sided Student's t-test. Vectors show the rotational eddy wind trend implied by the stream function trend (reference arrow 2 m s$^{-1}$ per 34 yr). Boxes mark the central Pacific upper-level cyclonic gyre couplet (ERA5 centers $\pm$10$^\circ$ latitude and $\pm$25$^\circ$ longitude, excluding 3$^\circ$ either side of the equator). (b) Strength of the Gill-type Rossby gyre couplet for each member, half the difference between the southern and northern box-mean trends (both cyclonic), in 10$^6$ m$^2$ s$^{-1}$ per 34 yr. Violins, black bars and whiskers are as in Figure~\ref{fig:jet_shift_scatter}, and the horizontal black line is ERA5. Filled dots indicate members with a statistically significant trend.}
    \label{fig:stream_function_trends}
\end{figure}

We next examine an additional circulation trend linked more closely to SST forcing, referred to as a Matsuno-Gill-type pattern and its extratropical teleconnections via bipolar propagating Rossby wave trains (Figure~\ref{fig:stream_function_trends}). This pattern is the most prominent signal of a forced response to prescribed historical SSTs when examining long-term trends over the satellite era~\cite{SongYeChiDaoJinkudenoZhunDiHengFengDeYunDong1966, gillSimpleSolutionsHeatInduced1980, ambrizziStationaryRossbywavePropagation1997b, baxterHowTropicalPacific2019}. To quantify the response of circulation to this SST forcing, we define a Rossby wave gyre couplet using the two cyclonic stream function cells on either side of the primary diabatic heating in the central tropical Pacific (black boxes in Figure~\ref{fig:stream_function_trends}a). The pattern is reproduced in both physics-based models (with and without direct CO$_2$ or aerosol forcing) and almost all of the AIWCMs (Figure S9). The spread across AIMIP is 1.7 times that across the physics-based models (1.5 times without MD-1p5; Figure~\ref{fig:stream_function_trends}b). The AIWCMs show a weaker response to the SST forcing than the AMIP ensemble, with ERA5 outside the 5$^{th}$-95$^{th}$ percentile of all AIWCM spreads. Even though this circulation response is strongly linked to SST forcing, the AIWCMs collectively underestimate the magnitude of the response, highlighting a second regional circulation deficiency in the AIWCMs.

These stream function trends link together temperature and circulation metrics examined in this study. Both the successes and the discrepancies in the vertical structure of tropical temperature trends are consistent with SST-driven convective processes and latent heating, which drive the Rossby wave patterns seen in Figure~\ref{fig:stream_function_trends}. AIWCMs are able to capture the pattern of the circulation response to tropical heating, though with a weaker magnitude compared with ERA5 and AMIP (Figure~\ref{fig:stream_function_trends}). In most AIWCMs, anticyclonic trends in the Southern Hemisphere midlatitudes are consistent with a poleward shift of the jet stream or weakening on the equatorward side and strengthening on the poleward side of the jet (Figure~\ref{fig:jet_shift_scatter}). More regionally, most AIWCMs and physics-based models struggle to capture the pattern and magnitude of satellite-era stream function trends over the North Atlantic and Europe, consistent with the underestimation of the meridional wind trends (Figure~\ref{fig:europe_vwinds_trends} \& S7-S9).

\section{Conclusions \& Discussion}

\subsection{Conclusions}
In this study, we examine the ability of state-of-the-art AIWCMs submitted to AIMIP to capture regional satellite-era thermodynamic and circulation trends. Forced only by observed SSTs and sea ice, the AIWCMs range from matching or exceeding physics-based models in satellite-era warming to producing almost none. Land and tropical upper-level temperature trends are matched or outperformed by the consistently highest performing AIWCMs. However, another set of AIWCMs shows relatively weak trends over land or in the upper tropical troposphere. Low-level tropical specific humidity trends are near-zero or increasing in the AIWCMs, with the strongest upper-level tropical warming coinciding with the strongest moistening. This relationship is consistent with the physics-based models, but ERA5, which the AIWCMs were trained on, shows a weak upper-level warming with a low-level drying. This need not indicate a failure of the AIWCMs and physics-based models, but it highlights potential uncertainties associated with tropical humidity trends. Across these thermodynamic metrics, the intermodel spread, a common gauge for the structural uncertainty across models, is 2--6 times larger for the AIMIP ensemble compared with the physics-based models (2--5 times without MD-1p5). 

The AIWCMs are also able to reproduce the patterns of observed regional circulation trends but generally underestimate their amplitude, even though they see these trends in their training data. Regional meridional wind trends are underestimated over the North Atlantic and Europe, where ERA5 upper-level wind and wave-activity flux trends are far outside the 5$^{th}$-95$^{th}$ percentile spreads of the AIWCMs and physics-based models (Figure~\ref{fig:europe_vwinds_trends}). AIWCMs are also able to capture the pattern of the Matsuno-Gill gyre couplet, an upper-level dynamical response to tropical SST trends, but with weaker magnitudes compared to ERA5 and physics-based models (Figure~\ref{fig:stream_function_trends}). Despite underestimating the strength of these trends, the AIWCMs generally agree much more closely than for the thermodynamic trends, with their spread decreasing to 0.8--1.7 times the physics-based model spreads (0.8--1.5 times without MD-1p5). 

When evaluating these trends, it is important to consider that ERA5 contains the full historical forced response. Therefore, agreement with the observed trends does not by itself show that the AIWCMs reproduce the physical relationships behind them. This is shown by stratospheric cooling in AIWCMs without CO$_2$ forcing (Figure~\ref{fig:jet_shift_scatter}e), where most of the AIWCMs cool despite not including CO$_2$ in their training or inference and struggle to produce QBO-like oscillations~\cite<Figure~\ref{fig:qbo_u50};>{baxterBenchmarkingAtmosphericCirculation2026}. Similarly, some AIWCMs, like ArchesWeatherGen-V2, capture the magnitude of the austral summer jet shift seen in ERA5 while not capturing the upper-level meridional temperature trends associated with it (Figure~\ref{fig:jet_shift_scatter}d). ACE2-ERA5 also captures the jet shift, but attributes a much stronger 250 hPa polar cooling to CO$_2$, though this signal has been partially attributed to ozone. As in~\citeA{ruckerBenchmarkingRegionalThermodynamic2026}, fixed CO$_2$ experiments highlight these attribution pitfalls of the data-driven emulators, where many of the trends that physics-based models attribute to SSTs are no longer captured without time-varying CO$_2$. 

\subsection{Discussion}
Across the AIMIP ensemble, the AIWCMs show a large range of performance. Across the temperature and moisture metrics, the 5$^{th}$-95$^{th}$ percentile range of the pooled AIMIP members is 2 to 6 (2 to 5 without MD-1p5) times wider than that of the physics-based models. For circulation metrics, the spread across AIWCMs is generally comparable to that in physics-based models (0.8 to 1.7 times, 0.8 to 1.5 without MD-1p5). The spreads in the trends within individual AIMIP models are somewhat larger than those in individual physics-based model ensembles. This may reflect variations in ensemble generation, where stochastic or generative AIWCMs require different approaches to creating their ensembles~\cite{scherEnsembleMethodsNeural2021, maheshHugeEnsemblesPart2025, hennAIMIPPhase12026} and is an important consideration for future benchmarking studies as more AI approaches are developed from stochastic architectures. The different choices in architecture and training are both important aspects of intermodel spread that can reflect structural uncertainty of the AIWCM architectures but also randomness in the training process~\cite{maheshHugeEnsemblesPart2025}. Therefore, the spread across AIWCMs is not directly comparable with the spread across physics-based models, but rather can be a reflection of the different choices in architecture and training. For example, ArchesWeather-V2 and its generative counterpart ArchesWeatherGen-V2 share a backbone, yet their tropical upper-tropospheric temperature trends differ by more than the 5$^{th}$-95$^{th}$ percentile range of the physics-based models. However, the spread across these ensembles is greatly reduced when accounting for abrupt jumps that occur between 1981 and 1994 (Figure~S10). Even with a common AIMIP protocol, the growing variety of emulator architectures requires additional consideration of ensemble generation to effectively compare with physics-based models and decompose epistemic, aleatoric, and training uncertainty~\cite{bulteUncertaintyQuantificationDataDriven2026}.

In this study, we examine whether AIWCMs are capturing the right trends for the right reasons~\cite{simpsonConfrontingEarthSystem2025a}. 
In some cases, there are satellite-era trends that are not captured by either AIWCMs or physics-based models. Neither the AIWCMs nor the physics-based models capture the tropical humidity trends or the meridional wind trends over Europe. The AIWCMs are trained on ERA5 and are exposed to some extent to these trends in their training data, though they are trained on global loss functions. In the case of the tropical humidity trends, the negative ERA5 tropical humidity trends are not consistent across reanalysis products, suggesting that the ERA5 trends may be an artifact of data assimilation~\cite{allanGlobalChangesWater2022, christensenErrorERA52m2026}. Although AIWCMs weakly follow trends in the training data, it is unclear from diagnostic analysis why they do not reproduce this signal from their training data. Benchmarking and training with multiple reanalyses and observational products can help separate model versus training data uncertainties, as in Figure~\ref{fig:zm_trends_hus}e~\cite{christensenErrorERA52m2026}. A similar collective underestimate occurs in the meridional wind trends over Europe, where both AIWCMs and physics-based models underestimate the strength of the wave train pattern and the eastward wave-activity flux. Studies have linked this underestimation to a weak response to anthropogenic aerosol forcing in physics-based models~\cite{kangSummertimeStationaryWave, roldan-gomezAcceleratedEuropeanSummer2026}, but this is not included in the AIMIP protocol. Emulators trained in the full ERA5 record that includes the aerosol response might have been expected to capture this signal but not. 

In contrast to missing trends, we also find examples of trends or variability where the AIWCMs are likely ``memorizing'' or overfitting to the training data. The clearest example is the QBO, where cBottle-1-3 reproduces the ERA5 QBO from 1980 to 2014 ($r=0.98$), but immediately when entering the testing period no longer correlates with ERA5 ($r=-0.1$, Figure~\ref{fig:qbo_u50}h). Diffusion-based methods have been suggested to struggle with generalization~\cite{guMemorizationDiffusionModels2023,carliniExtractingTrainingData2023, kadkhodaieGeneralizationDiffusionModels2023, brenowitzClimateBottleGenerative2025}. Because cBottle-1-3 generates each sample independently, conditioned only on monthly SST and the date~\cite{brenowitzClimateBottleGenerative2025}, it has no memory with which to propagate a QBO. Its in-phase QBO during training therefore indicates that it retrieves the training-period stratospheric state from the SST pattern, which fails once the SSTs leave the training record. Land temperature behaves differently because it is largely set by the SSTs. cBottle-1-3 retains, and even improves, its year-to-year relationship with ERA5 in the testing period (detrended $r=0.78$ in training vs. $r=0.87$ in testing), although it becomes colder ($-0.04$ K in training vs. $-0.85$ K in testing, Figure~\ref{fig:land_temperature_trends}d). The ability of diffusion-based sampling to generalize outside their training period may therefore depend on how strongly the variable is constrained by the forcings they are conditioned on. 

CO$_2$-ablation experiments with ACE2-ERA5 show that, for some trends, an AIWCM can also reproduce a trend through a different forcing pathway than that seen in physics-based models. For example, the temperature trends and poleward shift of the austral summer jet are captured in ACE2-ERA5 when given time-evolving CO$_2$ forcing, but physics-based models attribute these trends in large part to indirect SST forcing, consistent with analysis of temperature and land humidity trends in~\citeA{ruckerBenchmarkingRegionalThermodynamic2026}. 
In contrast, there are other examples, such as the North Atlantic and European meridional wind trends and Matsuno-Gill response to SSTs, where the AIWCMs underestimate the trends even with CO$_2$ included in training.
In~\citeA{clarkDisentanglingEffectsSea2026b}, a combination of training on physics-based model simulations and decoupling CO$_2$ from time in training has been shown to produce a more physically consistent partitioning between forcings in the near-surface temperature field, though this is not currently a viable path forward for AIMIP and AIWCMs trained on ERA5. 

The results of this study have implications for the use of AIWCMs in AMIP-style tasks and future projections. 
Proposed uses of emulators focus on creating larger ensembles than are currently feasible with physics-based models to probe uncertainty associated with internal variability and better sample the tails of the distribution for extreme events~\cite{eyringPushingFrontiersClimate2024, vankatwykRewiringClimateModeling2026, maheshHugeEnsemblesPart2025, mahajanStatisticalEquivalenceAI2026, duanAIbasedClimateModel2024, paciorekQuantifyingVeryExtreme2026, lancelinAIboostedRareEvent2025}. These uses require an AIWCM to translate the prescribed SSTs, sea ice and other forcings into the correct tropospheric response and to generate weather realizations conditioned on that response. The underestimated regional circulation trends, the thermodynamic and jet trends reproduced through CO$_2$ rather than SSTs, and the QBO that is reproduced only in one AIWCM during the training period show that current AIWCMs do not yet reliably capture this conditioning~\cite{ullrichRecommendationsComprehensiveIndependent2025, beuclerArtificialIntelligencePathways2026}.
This study also highlights several major uncertainties associated with current AIWCMs, including the spread across models and ensemble members, the ability to capture trends in temperature and circulation, and the attribution of these trends to different forcings. These uncertainties need to be carefully considered when using AIWCMs even for in-sample tasks, but especially for future projections or out-of-distribution cases~\cite{zhangEquilibriumResponseAtmospheric2026, hennAIMIPPhase12026}.

\section*{Open Research Section}
\begin{sloppypar}
ERA5~\cite{hersbachERA5GlobalReanalysis2020} is publicly available from the Copernicus Climate Change Service (C3S) Climate Data Store (CDS) as monthly averaged data on pressure levels (\mbox{\url{https://doi.org/10.24381/cds.6860a573}}) and on single levels (\mbox{\url{https://doi.org/10.24381/cds.f17050d7}}) and as hourly data on pressure levels (\mbox{\url{https://doi.org/10.24381/cds.bd0915c6}}). JRA-3Q~\cite{kosakaJRA3QReanalysis2024} is available from the NSF NCAR Geoscience Data Exchange (\mbox{\url{https://doi.org/10.5065/avtz-1h78}}). MERRA-2~\cite{gelaroModernEraRetrospectiveAnalysis2017} 6-hourly analyzed fields on pressure levels (M2I6NPANA) are available from the NASA Goddard Earth Sciences Data and Information Services Center (\mbox{\url{https://doi.org/10.5067/A7S6XP56VZWS}}). AMIP and AMIP-piForcing data are available from the Earth System Grid Federation (ESGF) at \mbox{\url{https://esgf-node.llnl.gov/projects/cmip6/}}. Scripts and checkpoints needed to run ACE2-ERA5 are available from the Allen Institute for AI at \mbox{\url{https://huggingface.co/allenai/ACE2-ERA5}}. Scripts and checkpoints needed to run NeuralGCM are available from \mbox{\url{https://github.com/neuralgcm/neuralgcm}}. AIMIP data are available from the AIMIP archive (\mbox{\url{https://github.com/ai2cm/AIMIP}}); the code repository below includes the script used to download them. The analysis and plotting code used in this study is available at \mbox{\url{https://github.com/itbaxter/AIMIP-satellite-era-benchmarking}} (Baxter et al., \citeyearNP{baxterAIMIPSatelliteeraBenchmarking2026}) and the processed outputs of the ACE2-ERA5 (full forcing and fixed CO$_2$) and 37-member NeuralGCM ensembles run for this study are archived at Zenodo (\mbox{\url{https://doi.org/10.5281/zenodo.24414264}}; Baxter et al., \citeyearNP{baxterAIMIPSatelliteeraBenchmarkingData2026}).
\end{sloppypar}

\section*{Conflict of Interest declaration}
The authors declare no conflicts of interest.

\acknowledgments
The authors acknowledge support from National Science Foundation grant GCP-2616141. IB and TAS are supported by the National Oceanic and Atmospheric Administration award NA23OAR4310597. TAS also acknowledges support from the Simons Foundation (grant SFI-MPS-T-MPS-00007353). PH and KR are supported by National Science Foundation grant and by Schmidt Sciences, LLC. Computational resources were provided by NSF ACCESS (ATM170020), NCAR’s CISL (URIC0009), and the University of Chicago Research Computing Center. This work also received support from the University of Chicago’s Data Science Institute and Institute of Climate and Sustainable Growth through the AI for Climate Initiative.

\clearpage

\bibliography{main}

\end{document}